\pdfoutput=1
\documentclass[amsmath,amssymb,aps,prd,preprint,superscriptaddress,nofootinbib]{revtex4-1}

\usepackage{latexsym, graphicx, amssymb, amsmath, mathrsfs, physics}
\usepackage{setspace, bm}
\usepackage{grffile}
\usepackage{ifthen}
\usepackage[svgnames]{xcolor}
\usepackage[
  colorlinks=true,
  unicode=true,
  pdfstartview=FitV,
  linkcolor=OrangeRed,
  citecolor=MediumSeaGreen,
  urlcolor=RoyalBlue,
  linktocpage=true,
  bookmarks=true,
  bookmarksnumbered=true,
  pdftitle={Complex Langevin study for attractively interacting 1d two-component Fermi gas with population imbalance},
  pdfauthor={Shoichiro Tsutsui}]{hyperref}

\newboolean{debugmode}
\setboolean{debugmode}{true}

\newcommand{\e}{\mathrm{e}}

\newcommand{\up}{\uparrow}
\newcommand{\dwn}{\downarrow}
\newcommand{\del}{\partial}

\newcommand{\calD}{\mathcal{D}}

\newcommand{\hc}{\hat{c}}
\newcommand{\hH}{\hat{H}}
\newcommand{\hK}{\hat{K}}
\newcommand{\hN}{\hat{N}}
\newcommand{\bg}{\bar{g}}
\newcommand{\bmu}{\bar{\mu}}
\newcommand{\bpsi}{\bar{\psi}}
\newcommand{\bphi}{\bar{\phi}}

\DeclareMathOperator{\Tprod}{\mathrm{T}}

\begin{document}
\preprint{RIKEN-QHP-495}

\title{Complex Langevin study for polarons in an attractively interacting one-dimensional two-component Fermi gas}

\author{Takahiro M. Doi}
\email[]{takahiro.doi@rcnp.osaka-u.ac.jp}
\affiliation{
	Research Center for Nuclear Physics (RCNP), Osaka University, 567-0047, Japan
}

\author{Hiroyuki Tajima}
\email[]{hiroyuki.tajima@phys.s.u-tokyo.ac.jp}
\affiliation{
Department of Physics, Graduate School of Science,
The University of Tokyo, Tokyo 113-0033, Japan
}

\author{Shoichiro Tsutsui}
\email[]{shoichiro.tsutsui@riken.jp}
\affiliation{
	Theoretical Research Division, Nishina Center, RIKEN, Wako, Saitama 351-0198, Japan
}

\begin{abstract}
We investigate a polaronic excitation
in a one-dimensional spin-1/2 Fermi gas with contact attractive interactions, 
using the complex Langevin method, which is a promising approach to evade a possible sign problem in quantum Monte Carlo simulations.
We found that the complex Langevin method works correctly
in a wide range of temperature, interaction strength,
and population imbalance.
The Fermi polaron energy extracted from the two-point imaginary Green's function is not sensitive to the temperature and the  impurity concentration
in the parameter region we considered.
Our results show a good agreement with the solution of the thermodynamic Bethe ansatz at zero temperature.
\end{abstract}


\maketitle

\section{Introduction}
The quantum Monte Carlo method~\cite{VONDERLINDEN199253,Pollet2012} 
is widely used in various fields of physics as a non-perturbative tool of analysis.
In a path integral formalism using Lagrangian,
a partition function is written in terms of the integral of the Boltzmann weight 
$\e^{-S}$ over field variables, where $S$ is an action.
When the action is a real-valued function,
the Boltzmann weight is regarded as a probability density function.
This ensures that quantum expectation values of physical observables 
can be estimated by importance sampling of the Boltzmann weight.
However, the positivity of the Boltzmann weight is violated 
in many physically interesting systems:
Hubbard model, finite density quantum chromodynamics (QCD), QCD with a $\theta$-term, matrix superstring models and any systems defined by Schwinger-Keldysh formalism which describes real-time dynamics, for instance \cite{PhysRevB.41.9301, PhysRevLett.94.170201,Muroya:2003qs,deForcrand:2010ys,Vicari:2008jw,Krauth:1998xh, PhysRevD.75.045007, PhysRevLett.117.081602}.
In these cases, 
the number of samples becomes exponentially large as the system size grows
in order to obtain statistically significant results.
In non-relativistic fermionic systems, 
a frequently used way to apply the quantum Monte Carlo method is 
introducing bosonic auxiliary field 
through the Hubbard-Stratonovich transformation~\cite{Blankenbecler:1981jt,Scalapino:1981ju,SUGIYAMA19861}.
After integrating out the fermion fields,
we will obtain an effective action of the auxiliary field.
Since the effective action involves a logarithm of a fermion determinant,
the positivity is not guaranteed except in a few cases
where the action has the particle-hole symmetry~\cite{PhysRevB.31.4403},
the Kramers symmetry~\cite{PhysRevB.71.155115},
or Majorana positivity~\cite{Li:2014tla,Li:2016gte,PhysRevLett.116.250601},
for instance.

A promising approach to evade the sign problem 
is the complex Langevin method~\cite{Klauder:1983sp,Parisi:1984cs},
which is an extension of the stochastic quantization to systems with complex-valued actions.
An advantage of this method is that it is scalable to the system size,
and thus, the computational cost is similar to 
the usual quantum Monte Carlo method without the sign problem. 
On the other hand, it is known that this method sometimes gives incorrect answers
even when the statistical average of a physical observable converges.
In the recent decade, 
a way to judge the reliability of the complex Langevin method is extensively 
studied~\cite{Aarts:2009uq,Aarts:2011ax,Nishimura:2015pba,Nagata:2015uga,Nagata:2016vkn,Salcedo:2016kyy,Aarts:2017vrv,Nagata:2018net,Scherzer:2018hid,Scherzer:2019lrh,Cai:2021nTV},
and proposed criteria which are able to compute in actual simulations 
using the boundary terms~\cite{Aarts:2009uq,Aarts:2011ax,Scherzer:2018hid,Scherzer:2019lrh} 
and the probability distribution of the drift term~\cite{Nishimura:2015pba,Nagata:2016vkn}.
While it is still difficult to predict when the complex Langevin method fails 
without performing numerical simulations,
we can eliminate wrongly convergent results thanks to these criteria.
In the context of cold fermionic atoms, 
the complex Langevin method is applied to 
rotating bosons~\cite{Hayata:2014kra},
polarized fermions~\cite{PhysRevD.98.054507, Rammelmuller:2018hnk, 10.21468/SciPostPhys.9.1.014, PhysRevA.103.043330},
unpolarized fermions with contact repulsive interactions~\cite{Loheac.PhysRevD.95.094502}
and mass imbalanced fermions~\cite{Rammelmuller.PhysRevD.96.094506}
to study ground state energy, thermodynamic quantities and Fulde-Ferrell-Larkin-Ovchinnikov-type pairings
(see also a recent review~\cite{Berger:2019odf}).

In this paper,
we consider a spatially one-dimensional spin-1/2 polarized fermions with contact attractive interactions which is known as the Gaudin-Yang model~\cite{RevModPhys.85.1633},
and compute the single-particle energy of spin-down fermions in a spin-up Fermi sea,
which is referred to as the Fermi polaron energy.
Recently, the single-particle excitation spectra of Fermi polarons were experimentally measured in higher dimensional atomic systems~\cite{PhysRevLett.102.230402,PhysRevLett.103.170402,Koschorreck2012,Cetina96,PhysRevLett.118.083602,PhysRevLett.125.133401,PhysRevX.10.041019,PhysRevA.103.053314}
(also see a recent review~\cite{Massignan_2014} for Fermi polarons).
While an analytic formula for the polaron energy in one dimension is obtained exactly at zero temperature 
based on the thermodynamic Bethe ansatz method~\cite{doi:10.1063/1.1704798},
no analytical solutions are known at finite temperature (note that the numerical results for finite-temperature properties of polarized gases within the thermodynamic Bethe ansatz were reported in Ref.~\cite{PhysRevA.94.031604}).
The one-dimensional Fermi polarons were studied
with several theoretical approaches such as Bruckner-Hartree-Fock~\cite{PhysRevA.84.033607}, {\it T}-matrix~\cite{PhysRevLett.111.025302,atoms9010018}, and variational~\cite{YSong2019,Mistakidis_2019} approaches.
In this study, 
we demonstrate that a microscopic quantity, that is, the polaron energy is efficiently computed by the complex Langevin method.

This paper is organized as follows.
In Sec.~\ref{sec:GY}, 
we derive a lattice action of the Gaugin-Yang model.
In Sec.~\ref{sec:CLM},
we review how to compute physical quantities using the complex Langevin method.
In Sec.~\ref{sec:Observables},
we show a way to extract the ground state energy in the spin-down channel 
from a two-point imaginary time Green's function.
In Sec.~\ref{sec:results},
we present the numerical results.
Section~\ref{sec:summary} is devoted to the summary of this paper.
In this work, $k_{\rm B}$ and $\hbar$ are taken to be unity.

\section{The Gaudin-Yang model}\label{sec:GY}
We consider a one-dimensional two-component Fermi gas 
with contact attractive interactions 
which is known as the Gaugin-Yang model~\cite{PhysRevLett.19.1312,GAUDIN196755}. 
The Hamiltonian is given by
\begin{align}
	\hat{H} 
	= 
	\sum_{p,\sigma}
	\qty(\frac{p^2}{2} - \mu_\sigma) 
	\hc_{p,\sigma}^\dag 
	\hc_{p,\sigma}
	-
	g
	\sum_{p,p',q} 
		\hc_{p+\frac{q}{2},\up}^\dag 
		\hc_{-p+\frac{q}{2},\dwn}^\dag 
		\hc_{-p'+\frac{q}{2},\dwn}
		\hc_{p'+\frac{q}{2},\up},
		\label{Hamiltonian}
\end{align}
where 
	$\hc_{p,\sigma}$ and $\hc_{p,\sigma}^{\dag}$ 
are fermionic annihilation/creation operators with 
momentum $p$ and spin $\sigma=\uparrow,\downarrow$, respectively.
In this work, the atomic mass is taken to be unity.
The coupling constant $g$ is related to 
a scattering length in one dimension $a$ as $g=\frac{2}{a} > 0$~\cite{RevModPhys.85.1633}. 
The chemical potential of spin-$\sigma$ fermions 
are represented by $\mu_\sigma$.
For convenience, we introduce an average chemical potential 
	$\mu=(\mu_{\up}+\mu_{\dwn})/2$ 
and a fictitious Zeeman field 
	$h=(\mu_{\up}-\mu_{\dwn})/2$. 
The grand canonical partition function is given by 
	$Z = \Tr\bqty{
		\e^{-\beta \qty(\hH-\sum_{\sigma}\mu_{\sigma}\hN_\sigma)}
	}$ 
with $\beta$ being an inverse temperature and a number operator 
	$\hN_\sigma = \sum_{p}\hc_{p,\sigma}^\dag \hc_{p,\sigma}$.
The path-integral representation of $Z$ reads
\begin{align}
	Z = \int \prod_{\sigma} \calD\psi^*_\sigma \calD\psi_\sigma \ \e^{-S},
\end{align}
where action $S$ is given by
\begin{align}
	S = 
	\int_0^\beta \dd{\tau} \int \dd{x}
	\Bigg[ 
		&\sum_{\sigma=\uparrow, \downarrow}
		\psi^*_{\sigma}(x,\tau) \qty(
			\frac{\del}{\del \tau} 
			- \frac{1}{2}\frac{\del^2}{\del x^2} 
			- \mu_\sigma
			)
		\psi_\sigma(x,\tau) \notag \\
		&\qquad\qquad 
		- g 
		\psi^*_\uparrow(x,\tau) 
		\psi^*_\downarrow(x,\tau) 
		\psi_\downarrow(x,\tau) 
		\psi_\uparrow(x,\tau)
	\Bigg].
	\label{action}
\end{align}
Here $\psi_\sigma(x,\tau), \psi^*_\sigma(x,\tau)$ 
are a Grassmann field and its complex conjugate.

While the action~\eqref{action} is given in a continuous spacetime, 
one should perform a lattice regularization appropriately 
to carry out numerical simulations.
We write lattice spacing of temporal and spatial directions
by $a_\tau$ and $a_x$, respectively, and their ratio by $r = a_\tau/a_x^2$.
We also introduce lattice quantities by
\begin{align}
	\bmu_\sigma \equiv \mu_\sigma a_x^2, \quad 
	\bg \equiv g a_x, \quad 
	\bpsi_{\sigma,j,n} \equiv \psi_\sigma(ja_x,na_\tau) a_x^{1/2},
\end{align}
where $n$ and $j$ are integers that satisfy 
$0 \leq n < N_\tau$ and $0 \leq j < N_x$.
The inverse temperature and the spatial length of the lattice is 
given by $\beta = T^{-1} = N_\tau a_\tau$ and $L = N_x a_x$.
With these notations, we consider a lattice action:
\begin{align}
	S_\text{lat}
	=
	&\sum_{j,n} \sum_{\sigma=\uparrow,\downarrow} 
	\qty(
		\bar{\psi}^*_{\sigma;j,n} \bar{\psi}_{\sigma;j,n} 
		- 
		\bar{\psi}^*_{\sigma;j,n+1} 
		e^{-\bar{\phi}_{j,n} + \bar{\mu}_\sigma} 
		\bar{\psi}_{\sigma;j,n} 
		+ 
		\frac{r}{2} 
		(\bar{\psi}^*_{\sigma;j+1,n} - \bar{\psi}^*_{\sigma;j,n}) 
		(\bar{\psi}_{\sigma;j+1,n} - \bar{\psi}_{\sigma;j,n})  
	) \notag \\
	&+\sum_{j,n} \frac{\cosh(\bar{\phi}_{j,n}) - 1}{\bar{g}},
	\label{lattice action}
\end{align}
where $\bar{\phi}_{j,n}$ is a bosonic auxiliary field.
As shown in Ref.~\cite{Alexandru:2018brw}, 
the lattice action~\eqref{lattice action} correctly converges to the continuum one  
as long as the matching conditions
\begin{align}
	\frac{g a_\tau}{a_x} 
	= 
	\qty(\frac{f_2}{f_0} - \frac{f_1^2}{f_0^2}) 
	\e^{\bar{\mu}_\uparrow + \bar{\mu}_\downarrow},
	\quad
	\mu_\sigma a_\tau 
	= \frac{f_1}{f_0} \e^{\bar{\mu}_\sigma} - 1 
\end{align}
are satisfied, where $f_k$ is a $\bg$-dependent constant given by
\begin{align}
	f_k 
	= 
	\int_{-\infty}^\infty \dd{\bar{\phi}}
	\e^{-\frac{\cosh(\bar{\phi}) - 1}{\bar{g}}} \e^{k\bar{\phi}}.
\end{align}
In practice, it is sufficient to use an approximated form of the matching conditions
\begin{align}
	\bar{g} \simeq \frac{ga_\tau}{a_x}, \quad 
	\bar{\mu}_\sigma \simeq \mu_\sigma a_\tau - \frac{ga_\tau}{2a_x},
\end{align}
which are obtained as the first order approximation 
in the expansion in terms of $a_\tau$.
After integrating out the fermion fields, 
the partition function and the effective action of the auxiliary field read
\begin{align}
	Z 
	&= 
	\int \prod_{j,n} \dd{\bar{\phi}_{j,n}} 
	\e^{-S_\text{eff}[\bar{\phi}]},
	\label{partitionfunction}
\end{align}
where the effective action of the auxiliary field is given by
\begin{align}
	S_\text{eff}[\bar{\phi}]
	&=
	\sum_{j,n} \frac{\cosh\bar{\phi}_{j,n} - 1}{\bar{g}}
	- \sum_{\sigma} \log \det 
	\qty[ 
		I
		+ 
		e^{N_\tau \bar{\mu}_\sigma}
		B^{-1}C_{N_\tau-1} 
		\cdots 
		B^{-1}C_{0}
	], \label{effective action}\\
	B_{j,j'} 
	&= 
	-\frac{r}{2}
	\qty(
		\delta_{j-1,j'} + \delta_{j+1,j'}
	)
	+
	\qty(
		1 + r
	)
	\delta_{j,j'}, \quad
	(C_n)_{j,j'}
	=
	\delta_{j,j'} e^{-\bar{\phi}_{j,n}},
\end{align}
where $I$ is the $N_x \times N_x$ identity matrix.
Since we consider a naive finite difference as 
an approximation of the second order derivative with respect to $x$,
the eigenvalues of $B$ are 
$1+2r \sin^2\frac{\pi k}{N_x}, (k=0,1,\cdots, N_x-1)$.
It has been argued in Ref.~\cite{Blankenbecler:1981jt, Alexandru:2018brw} that
this naive lattice action converges too slow to the continuum limit,
and the behavior can be improved by replacing the eigenvalues of $B$ by 
\begin{align}
	\lambda_k = \exp(\frac{r}{2} \qty(\frac{2\pi k}{N_x})^2). 
\end{align}
After this replacement, the form of $B_\sigma$ is given by
\begin{align}
	B_{j,j'}
	=
	\frac{1}{N_x}
	\sum_{k=-\lfloor N_x/2 \rfloor}^{\lfloor N_x/2 \rfloor}
	\lambda_k
	\cos k(j-j').
\end{align}
A notable point is that 
the effective action~\eqref{effective action} 
involves a logarithm of the fermion determinant
which can be complex in general. 
Therefore, this term may cause the sign problem if the Zeeman field $h$ is not zero 
and then the Monte Carlo simulation can be
difficult to apply to this system.

\section{Complex Langevin method}\label{sec:CLM}
The complex Langevin method (CLM)~\cite{Klauder:1983sp,Parisi:1984cs} 
is an extension of the stochastic quantization 
which is usually applicable to real-valued actions.
In the CLM, we first consider a complexified auxiliary field 
	$\bphi_{n,k}$ 
and extend the domain of definition of $S_\text{eff}$ to the complex space.
For such a complex field, 
we consider a fictitious time evolution described by 
the complex Langevin equation:
\begin{align}
	\bphi^\eta_{n,k}(t + \Delta t) = 
	\bphi^\eta_{n,k}(t) 
	- \frac{\del S_\text{eff}}{\del \bphi^\eta_{n,k}} \Delta t 
	+ \eta_{n,k}(t) \sqrt{\Delta t},
\end{align}
where $\eta_{n,k}(t)$ is a real Gaussian noise.
When we assume that the system described by 
the complex Langevin equation reach equilibrium at $t = t_\text{eq}$,
an average of a physical observable $O(\bphi)$ can be defined as
\begin{align}
	\expval{O(\bphi)}
	\equiv 
	\lim_{T\to\infty}\frac{1}{T} 
	\int_{t_\text{eq}}^{t_\text{eq} + T} 
	\dd{t} 
	\expval{O(\bphi^\eta(t))}_\eta,
	\label{CLMaverage}
\end{align}
with the average over the noise 
	$\expval{O(\bphi^\eta(t))}_\eta$ 
being
\begin{align}
	\expval{O(\bphi^\eta(t))}_\eta 
	\equiv
	\frac{
		\int 
		\prod_{n,k,t} 
		\dd{\eta}_{n,k}(t) 
		O(\bphi^\eta(t)) 
		\e^{-\frac{1}{4} \sum_{n,k,t} \eta_{n,k}(t)^2}
	}{
		\int 
		\prod_{n,k,t} 
		\dd{\eta}_{n,k}(t) 
		\e^{-\frac{1}{4} \sum_{n,k,t} \eta_{n,k}(t)^2}
	}.
\end{align}
We note that 
	$\expval{\eta_{n,k}(t) \eta_{n',k'}(t')}_\eta 
	= 
	2\delta_{nn'}\delta_{kk'}\delta_{tt'}$, 
in particular.
Although, we expect that 
the mean value
	$\expval{O(\bphi)}$ 
is equivalent to the quantum expectation value calculated in an original action, i.e., 
	$\int \prod_{n,k} \dd{\bphi}_{n,k} 
	O(\bphi) \e^{-S_\text{eff}}
	/
	\int \prod_{n,k} \dd{\bphi}_{n,k} 
	\e^{-S_\text{eff}}$
in the limit $\Delta t \to 0$,
it is not correct in general.
There are extensive studies~\cite{Aarts:2009uq,Aarts:2011ax,Nishimura:2015pba,Nagata:2015uga,Nagata:2016vkn,Salcedo:2016kyy,Aarts:2017vrv,Nagata:2018net,Scherzer:2018hid,Scherzer:2019lrh,Cai:2021nTV} 
to understand when the CLM is justified,
and criteria for determining whether a CLM is reliable or not have been proposed.
One of a practical criterion 
which can be relied on in actual numerical simulations 
is discussed from a view point of a probability distribution of a drift term~\cite{Nishimura:2015pba,Nagata:2016vkn}.
In our case, 
it is sufficient to consider a magnitude of the drift term given by
\begin{align}
	v^\eta \equiv 
	\max_{n,k} \qty| 
		\frac{\del S_\text{eff}}{\del \bphi^\eta_{n,k}}
	| \label{drift_magnitude}
\end{align}
and its distribution.
According to the criterion, the CLM is reliable 
if the probability distribution of $v^\eta$ shows an exponential decay.

\section{Observables}\label{sec:Observables}
The number density of spin-$\sigma$ fermions is given by
\begin{align}
	n_\sigma 
	= 
	\frac{T}{L}\frac{\del}{\del \mu_\sigma} \log Z
	= 
	\frac{1}{L}
	\frac{1}{Z}
	\int \prod_{j,n} \dd{\bar{\phi}_{j,n}} 
	\tr \qty[
		\frac{1}
		{
			I + e^{-N_\tau \bar{\mu}_\sigma}
			C_{0}^{-1}B
			\cdots 
			C_{N_\tau-1}^{-1}B
		}
	]
	\e^{-S_\text{eff}[\bar{\phi}]}.
\end{align}
The particle number density on a lattice unit is defined by 
$\bar{n}_\sigma = n_\sigma a_x$.
From below,  
we assume that the spin-down fermions are regarded as minority.
Typical temperature and momentum scales are given by the Fermi scales 
which are determined by the density of spin-up fermions:
\begin{align}
	T_\text{F} = \frac{\pi^2 n_\up}{2}, 
	\quad 
	p_\text{F} = \pi n_\up.
\end{align}
In lattice simulations, 
we can compute dimensionless combination $T/T_\text{F}$ and $p_\text{F}a$ as follows:
\begin{align}
	\frac{T}{T_\text{F}} = \frac{2}{\pi^2 \bar{n}_\up^2 N_\tau r}, 
	\quad 
	p_\text{F}a = \frac{2\pi r \bar{n}_\up}{\bar{g}}. 
\end{align}

In order to calculate the polaron energy,
we consider two-point Green's function:
\begin{align}
	G(p,\tau) 
	\equiv 
	\frac{1}{Z}
	\Tr[ 
		e^{-\beta\hK} 
		\Tprod_\tau 
		\qty(
			\hc^\dagger_{\downarrow,p}(\tau) 
			\hc_{\downarrow,0}(0)
			) 
		],
	\quad 
	(-\beta \leq \tau \leq \beta),
\end{align}
where $\Tprod_\tau$ is the imaginary-time-ordered product.
Hereinafter we restrict $\tau > 0$.
We write the eigenvalue and eigenstate of $\hK$ by 
	$\hK\ket{n} = K_n\ket{n}$.
In particular, $K_0 < K_1 < \cdots$.
We also assume that the ground state $\ket{0}$ is not degenerated.
Expanding the trace by the eigenstates,
the correlation function reads
\begin{align}
	G(p,\tau) 
	=
	\frac{
	\sum_{nm}
		e^{-(\beta-\tau) \Delta K_n - \tau \Delta K_m}
		\mel{n}{\hc^\dagger_{\sigma,p}}{m}
		\mel{m}{\hc_{\sigma',p'}}{n}
	}{
		\sum_n 		
		e^{-\beta \Delta K_n}
	},
\end{align}
where $\Delta K_n \equiv K_n - K_0$.
In the low temperature limit $\beta \to \infty$,
only the ground state contributes to the summation over $n$.
Thus, we find
\begin{align}
	\tilde{G}(p,\tau)
	\equiv
	\lim_{\beta\to\infty}
	G(p,\tau) 
	=
	\sum_m
		e^{-\tau \Delta K_m}
		\mel{0}{\hc^\dagger_{\sigma,p}}{m}
		\mel{m}{\hc_{\sigma',p'}}{0}.
\end{align}
Since the matrix elements appeared in the above expression 
does not depend on $\tau$,
it behaves like
\begin{align}
	\tilde{G}(p,\tau)
	=
	A_0 e^{-\tau E_0} + A_1 e^{-\tau E_1} + \dots,
\end{align}
where $A_0, A_1, \dots$ are $\tau$-independent constants,
and $E_0, E_1, \dots$ are energies of the ground state and excited states.
In particular, the energy of the ground state can be extracted by
\begin{align}
	E_0(p)
	=
	\frac{1}{a_\tau}
	\lim_{\tau \to \infty} R(p,\tau),
	\quad 
	R(p,\tau) \equiv
	\log
	\frac{\tilde{G}(p,\tau)}
	{\tilde{G}(p,\tau+a_\tau)}, \label{ratio of green functions}
\end{align}
keeping $\tau \ll \beta$.
The polaron energy $U$ is defined by 
\begin{align}
U \equiv E_0(0) + \mu_\dwn. \label{def_polaron_energy}
\end{align}
The polaron energy is the shift of single-particle energy from that in the case of free fermions
due to the interaction between majority (spin-up fermions) and minority (spin-down fermions). 
We note that
the polaron energy at zero temperature is calculated exactly 
based on thermodynamic Bethe ansatz~\cite{doi:10.1063/1.1704798}:
\begin{align}
	\frac{U}{T_\text{F}}
	=
	-\frac{2}{\pi}\qty[
		\frac{1}{p_\text{F}a}
		+ \tan^{-1}\qty(\frac{1}{p_\text{F}a})
		+ \qty(
			\frac{\pi}{2} + \tan^{-1}\qty(\frac{1}{p_\text{F}a})
		) \frac{1}{(p_\text{F}a)^2}
	].
	\label{exact polaron energy}
\end{align}

In lattice calculations, 
the polaron energy is obtained as follows.
From the form of the effective action, 
the lattice expression of the inverse Green's function reads
\begin{align}
	G^{-1}
	\equiv 
	\begin{pmatrix}
		B & 0 & 0 & \cdots & 0 & \e^{\bar{\mu}_\dwn}C_{N_\tau-1} \\
		-\e^{\bar{\mu}_\dwn}C_0 & B & 0 & \cdots & 0 & 0 \\
		0 & -\e^{\bar{\mu}_\dwn}C_1 & B & \cdots & 0 & 0 \\
		\vdots  & \vdots  & \vdots & \ddots  & \vdots & \vdots \\
		0 & 0 & 0 & \cdots & -\e^{\bar{\mu}_\dwn}C_{N_\tau-2} & B
	\end{pmatrix}.
	\label{inverseGreenfunction}
\end{align}
From a straightforward algebra, 
each component of $G$ reads
\begin{alignat}{2}
	G_{jj} 
	&= \frac{B^{-1}}
	{
		I 
		+ 
		\e^{N_\tau\bar{\mu}_\dwn}
		B^{-1}C_{j-1} 
		\cdots 
		B^{-1}C_{0} 
		B^{-1}C_{N_\tau-1} 
		\cdots 
		B^{-1}C_j
	}, \\
	G_{kj} 
	&= 
	\begin{cases}
		B^{-1}C_{k-1} \cdots B^{-1}C_jG_{jj}, \quad &(j+1 \leq k \leq N-1), \\	
		-B^{-1}C_{k-1} \cdots B^{-1}C_0 B^{-1}C_{N-1} \cdots B^{-1}C_jG_{jj}, \quad &(0 \leq k \leq j-1).
	\end{cases}
\end{alignat}
The momentum representation of $G_{ij}$ is calculated by the discrete Fourier transformation.
Therefore, if the temporal lattice size $N_\tau$ is sufficiently large, 
$\tilde{G}(p,\tau)$ is approximately given by 
\begin{align}
	\tilde{G}(2\pi k/N_x, na_\tau)
	\simeq
	\frac{1}{N_x} \sum_{k',l'=0}^{N_x-1} 
	\e^{\frac{2\pi i}{N_x}kk'}
	(G_{0n})_{k'l'}.
\end{align}

\section{Numerical results}\label{sec:results}
We performed complex Langevin simulations on $(N_\tau, N_x) = (40, 60), (80, 60)$ lattices.
The anisotropy is set to $r = 0.1$.
There are three dimensionless parameters 
to characterize the Gaugin-Yang model: 
$\beta \mu$, $\beta h$ and $\lambda \equiv \sqrt{\beta} g$.
We fixed the dimensionless coupling constant by $\lambda = 2$, 
and swept the average chemical potential and the Zeeman field between 
$-1.2 \leq \beta \mu \leq 1.2, \, 0 \leq \beta h \leq 12$ for $N_\tau = 40$
and $-2.4 \leq \beta \mu \leq 2.4, \, 0 \leq \beta h \leq 24$ for $N_\tau = 80$,
respectively.
We set Langevin step-size by $\Delta t = 0.01$, 
and saved configurations of the auxiliary field at the 0.02 interval.
For every parameter sets, we took 5001 samples.
Error bars shown below are $1\sigma$ statistical errors 
calculated by the Jackknife method,
where bin-sizes are $0.3 - 1.2$ in units of Langevein time
depending on parameters and observables.

\begin{figure}[hbt]
	\centering
	\includegraphics[width=0.75\linewidth]{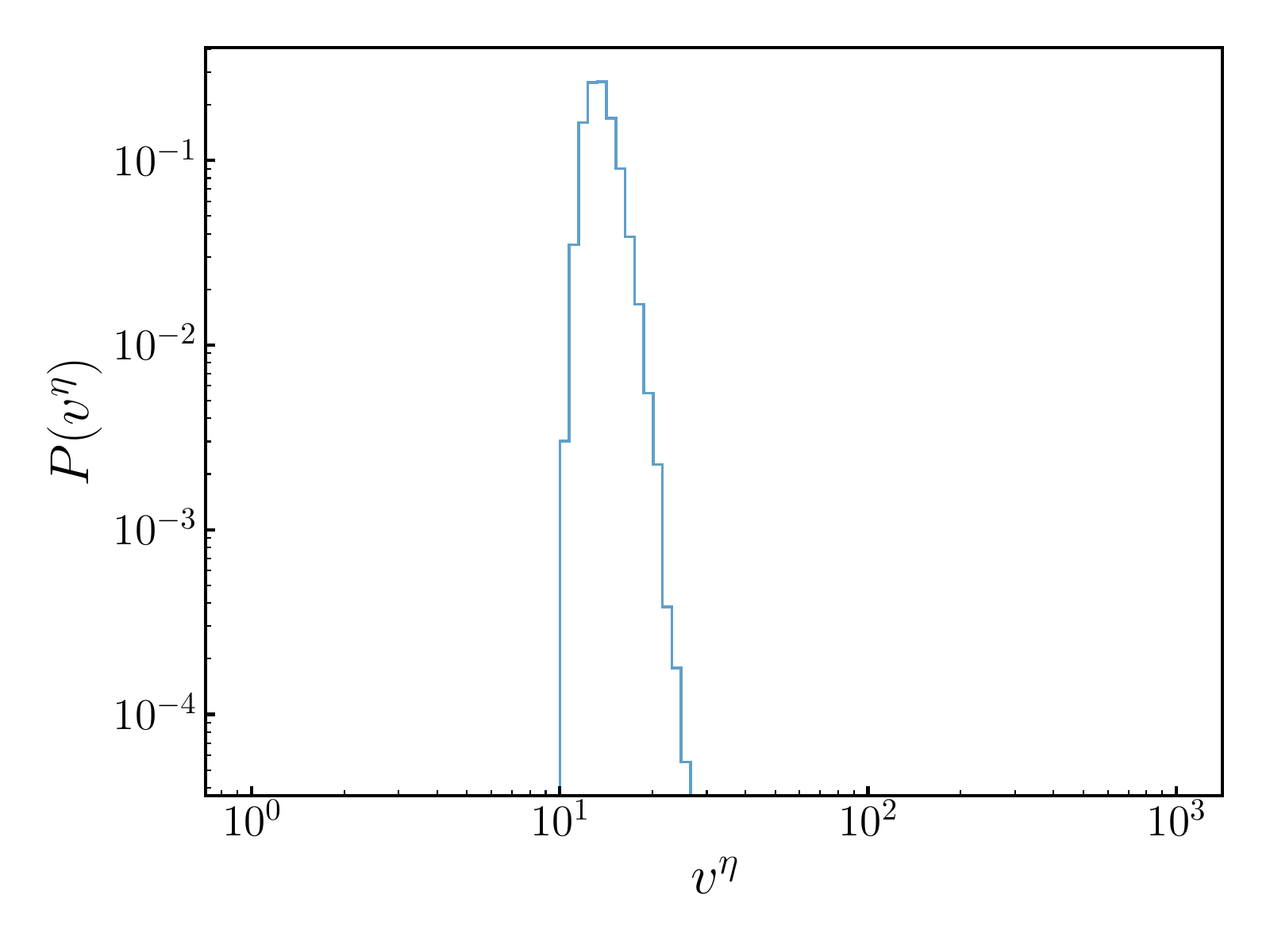}
	\caption{The histogram of the drift term for the $N_\tau=40$ lattice 
	at $\beta\mu=0$, $\beta h = 3$.}
	\label{fig:drift}
\end{figure}

In every Langevin step, 
the magnitude of the drift term (\ref{drift_magnitude})
is calculated and stored, 
and finally the probability distribution $P(v^\eta)$ of drift term
can be drown. 
In Fig.~\ref{fig:drift}, 
a typical result of the probability distribution $P(v^\eta)$ is shown. 
It is normalized so that the integral of the distribution is 1. 
In each simulation, 
we confirmed that $P(v^\eta)$ shows 
a linear fall-off in the log-log plot. 
This means that $P(v^\eta)$ shows an exponential fall-off, 
and then our calculation of CLM was reliable. 

\begin{figure}[hbt]
	\centering
	\includegraphics[width=0.75\linewidth]{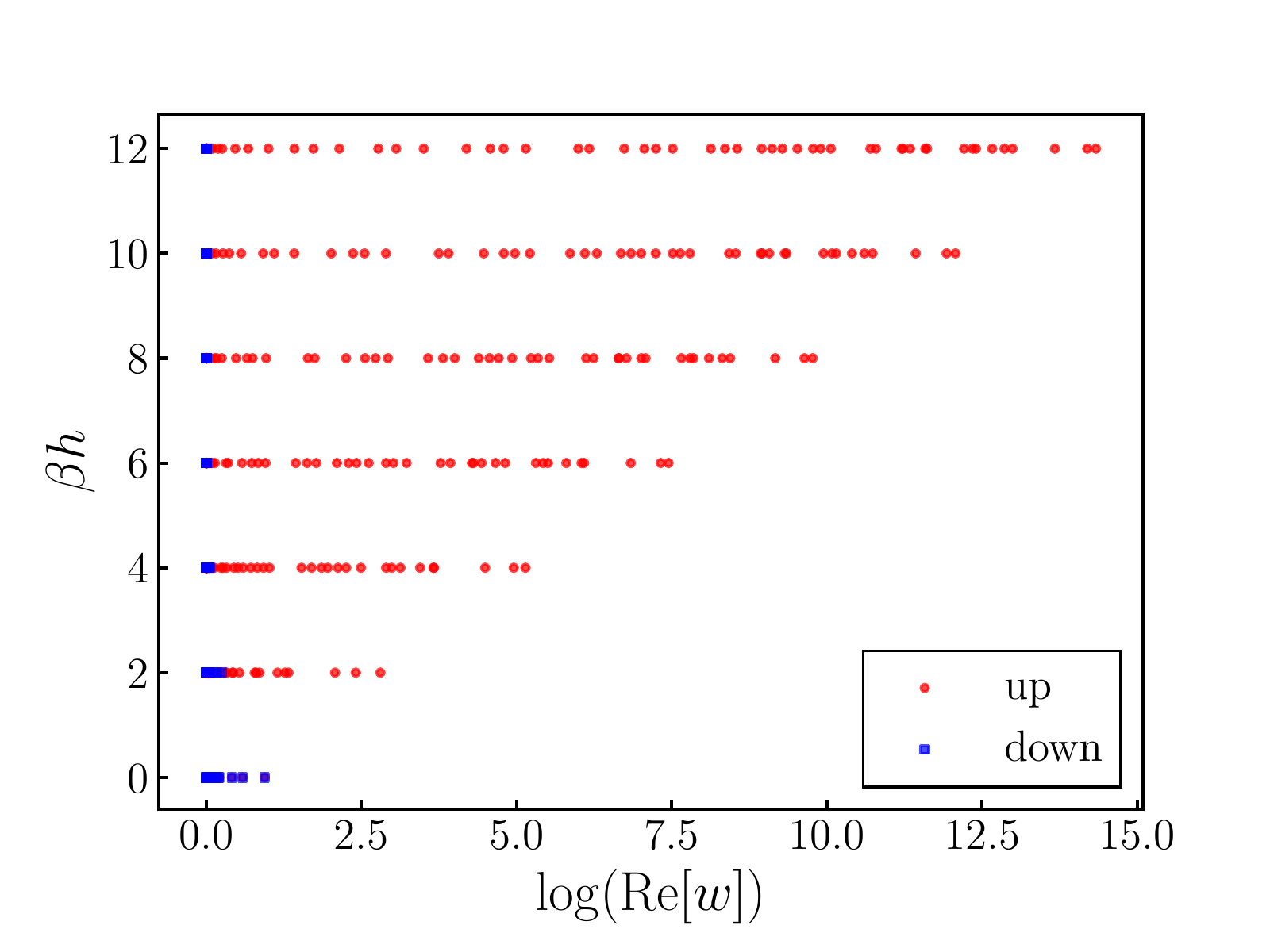}
	\caption{The eigenvalues of matrix $G^{-1}_{\sigma, {\rm red}}$ with several values of $h$ in the case of 
	$N_\tau=40$ and $\beta\mu=-1.2$. 
	Red circles and blue squares correspond to $G^{-1}_{\uparrow, {\rm red}}$ and $G^{-1}_{\downarrow, {\rm red}}$, respectively.
	}
	\label{fig:eigenvalue}
\end{figure}

We also investigated the eigenvalues 
of the matrix
\begin{align}
G^{-1}_{\sigma, {\rm red}}
=
I
+ 
e^{N_\tau \bar{\mu}_\sigma}
B^{-1}C_{N_\tau-1} 
\cdots 
B^{-1}C_{0},
\label{reducedinversGreenfunction}
\end{align}
which is the reduced matrix of inverse Green's function
Eq. (\ref{inverseGreenfunction}) 
appearing in the effective action on the lattice
Eq. (\ref{effective action}) 
as an effective fermionic matrix. 
We calculate the eigenvalues 
$w_\sigma$ of the matrix 
from one configuration 
in the case of several values of $\beta h=0$ to $\beta h=12$
and other fixed parameters, $N_\tau=40$ and $\beta\mu=-1.2$. 
The imaginary part of $w_\sigma$ is negligibly small
and hereinafter we discuss the real part of the 
eigenvalues. 
The numerical results of eigenvalues 
$\log {\rm Re}[w_\sigma]$
of the matrices 
$G^{-1}_{\uparrow, {\rm red}}$ and 
$G^{-1}_{\downarrow, {\rm red}}$ 
are shown 
in Fig. \ref{fig:eigenvalue}. 
Red circles and blue squares correspond to the eigenvalues of $G^{-1}_{\uparrow, {\rm red}}$ and $G^{-1}_{\downarrow, {\rm red}}$, respectively. 
In the case of $\beta h=0$, 
$w_\uparrow$ is exactly same as $w_\downarrow$ because 
$G^{-1}_{\uparrow, {\rm red}}= G^{-1}_{\downarrow, {\rm red}}$. 
While the
range of the eigenvalues of $G^{-1}_{\uparrow, {\rm red}}$ tend to be broad, 
the range of the eigenvalues of $G^{-1}_{\downarrow, {\rm red}}$ tend to be narrow 
when $\beta h$ increases. 

It is notable point of this eigenvalue-analysis 
that the eigenvalues of $G^{-1}_{\sigma, {\rm red}}$ 
are always larger than 1 because of $\log({\rm Re}[w])>0$ 
even in the case of the large $\beta h$, corresponding to the large population imbalance. 
This result indicates that the integrand of the partition function (\ref{partitionfunction}) is always positive and no sign problem happens in the parameter region of our calculations. 
Note that this is a numerical finding in our setup, 
and we do not prove that
the sign problem never occurs in the Gaudin-Yang model with population imbalance. 
We note that the sign problem may occur in other situations within the Hamiltonian~\eqref{Hamiltonian} or the action~\eqref{action}, for example, considering other values of masses, chemical potentials, coupling constants, lattice parameters, and dimensions. 

\begin{figure}[hbt]
	\centering
	\includegraphics[width=0.45\linewidth]{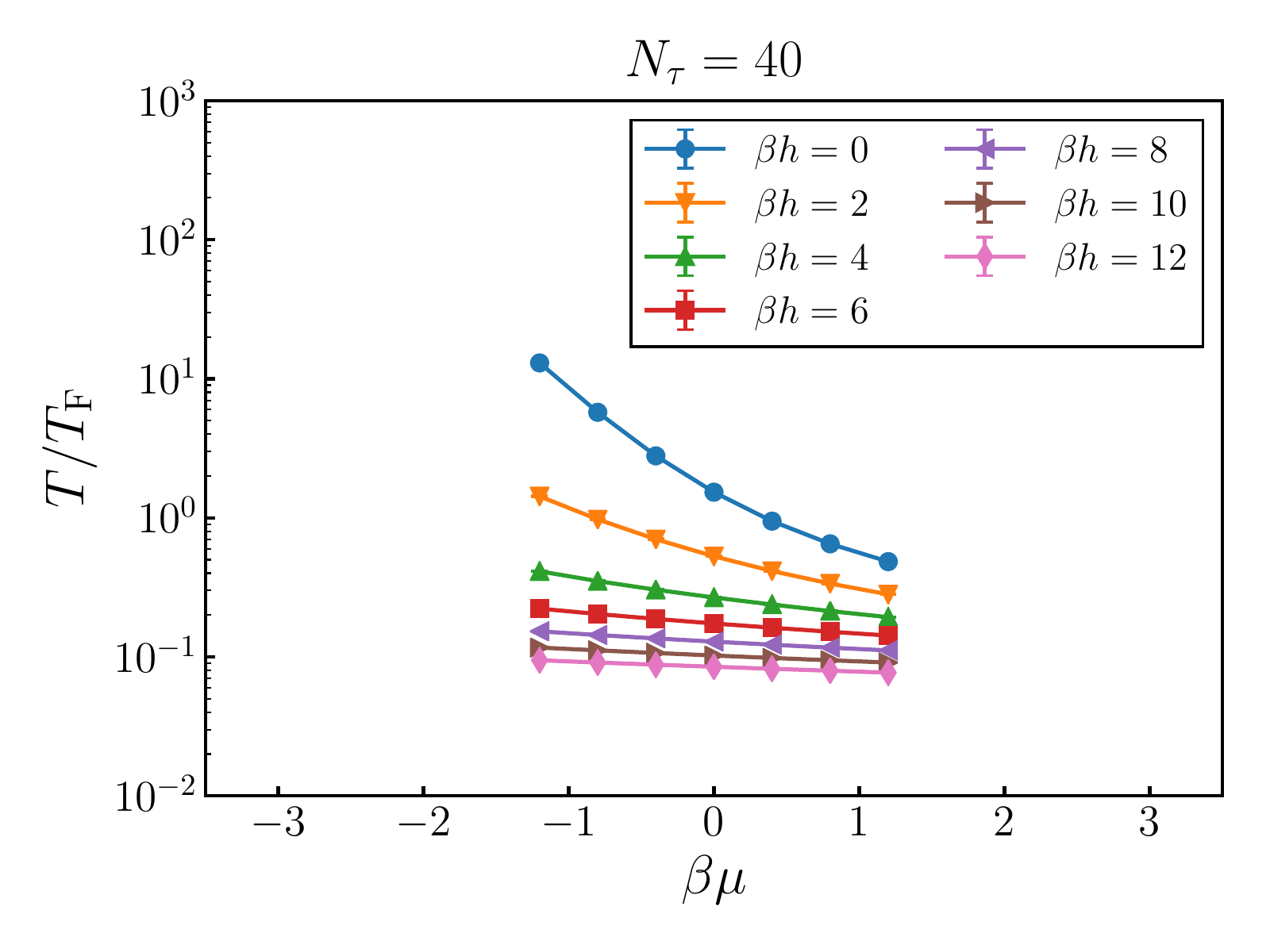}
	\includegraphics[width=0.45\linewidth]{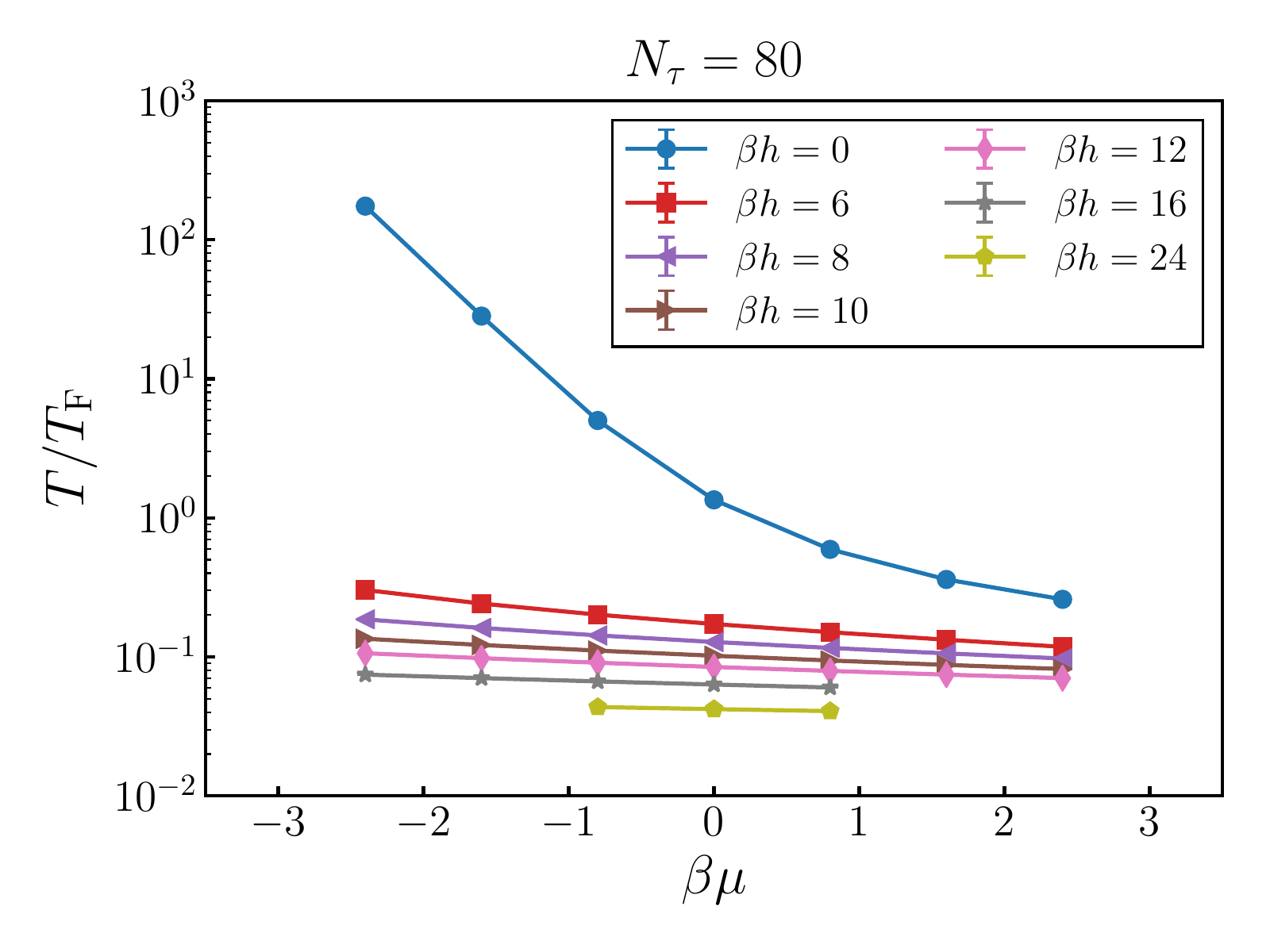}
	\includegraphics[width=0.45\linewidth]{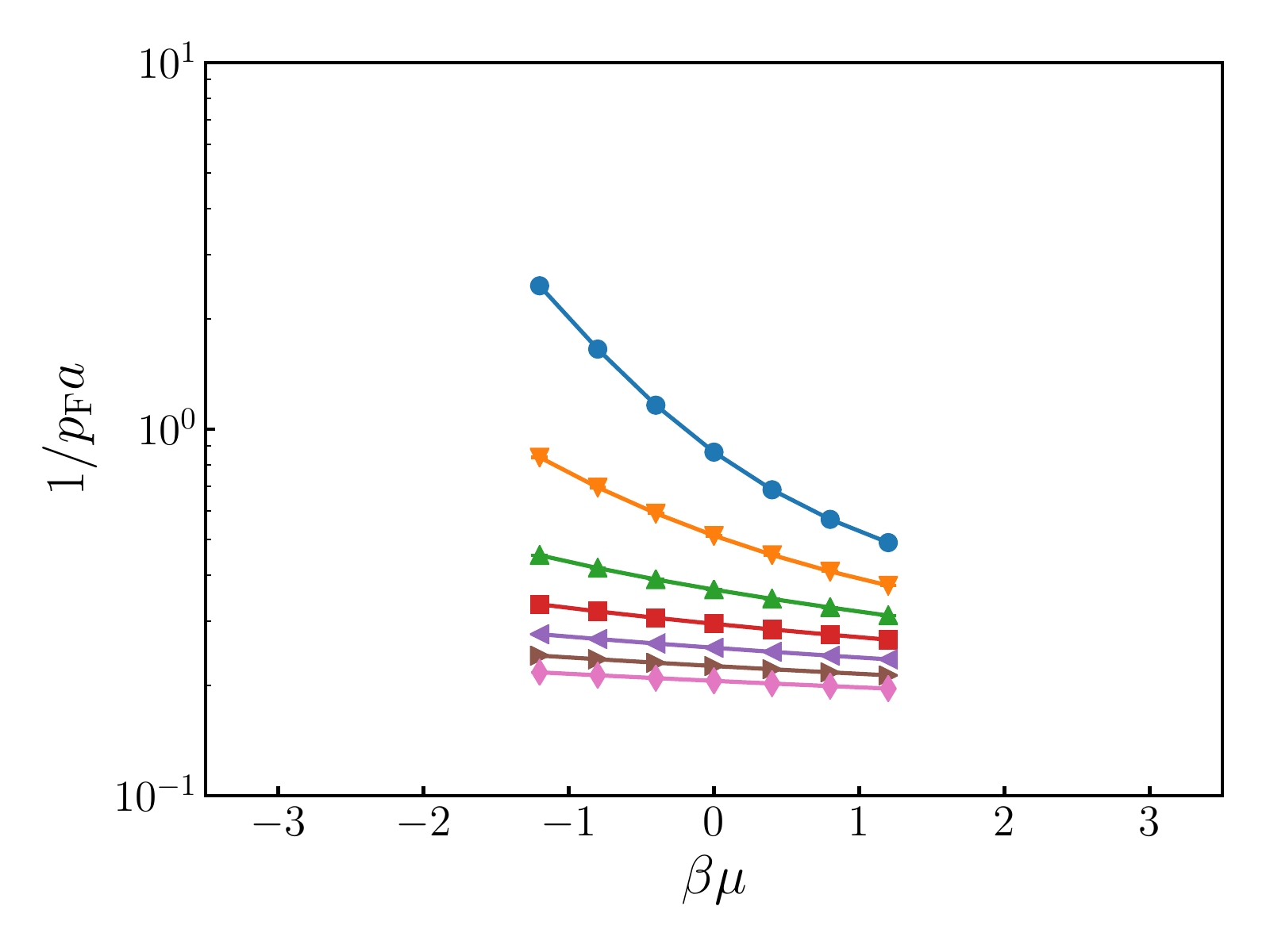}
	\includegraphics[width=0.45\linewidth]{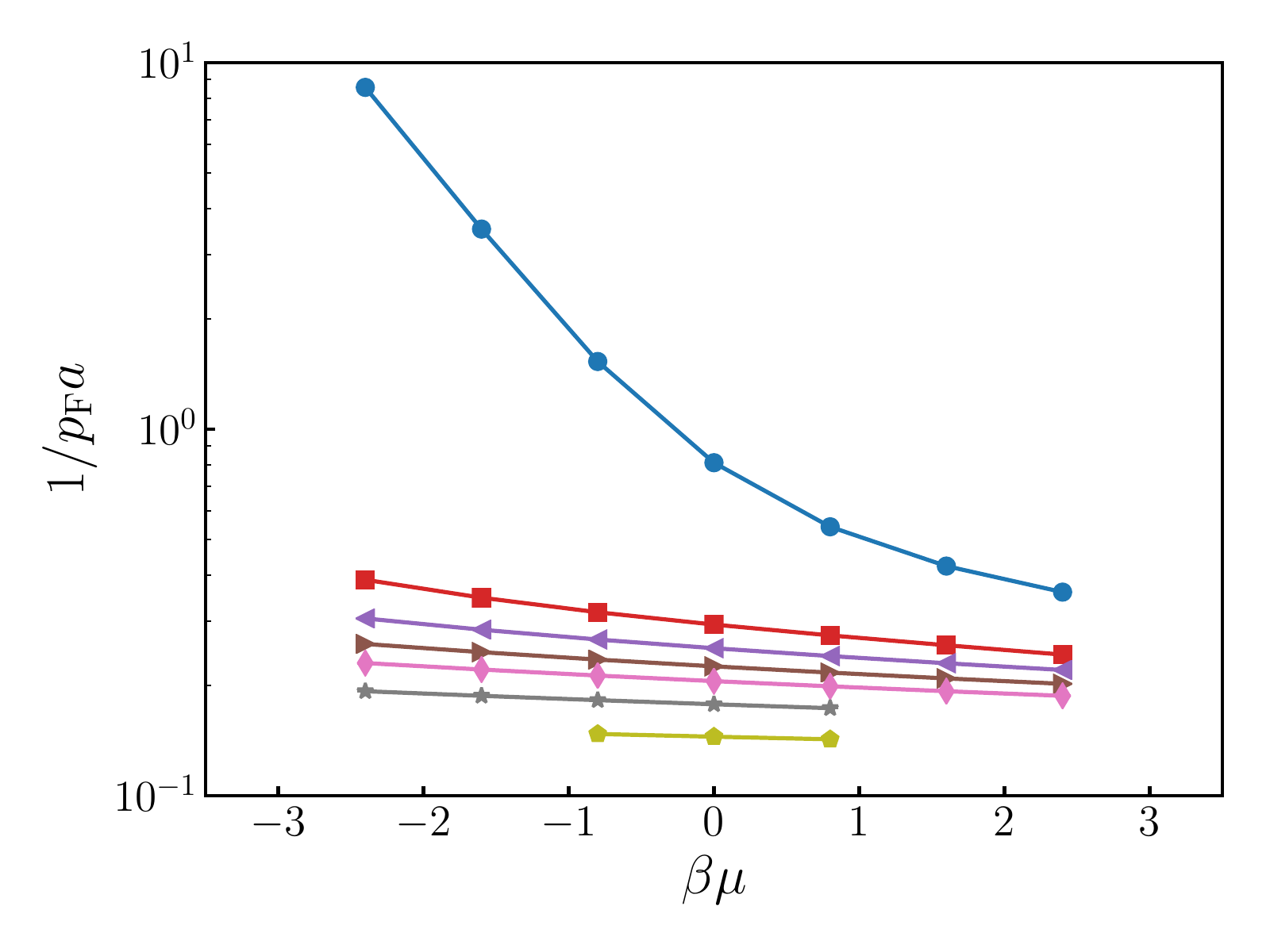}
	\includegraphics[width=0.45\linewidth]{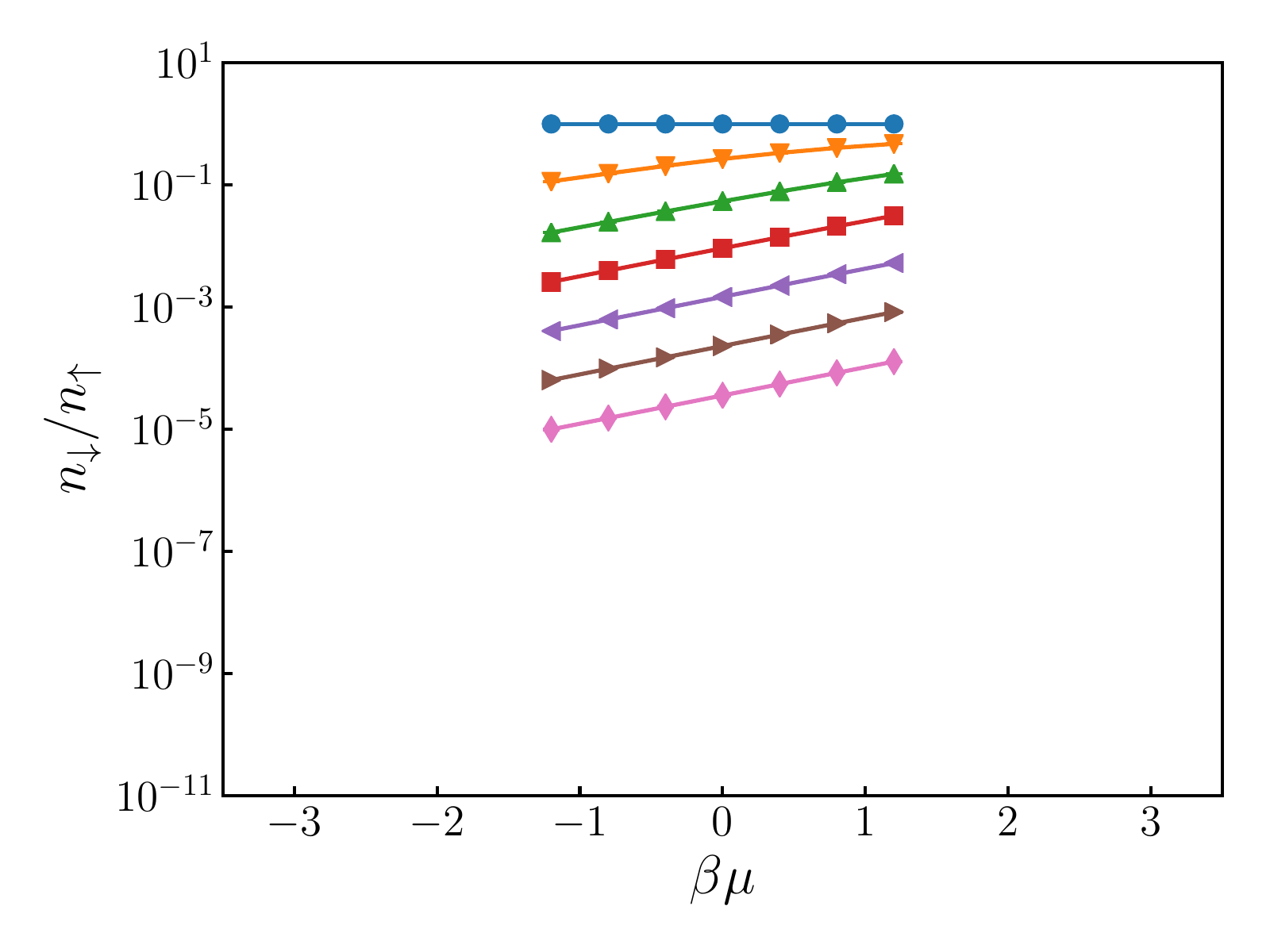}
	\includegraphics[width=0.45\linewidth]{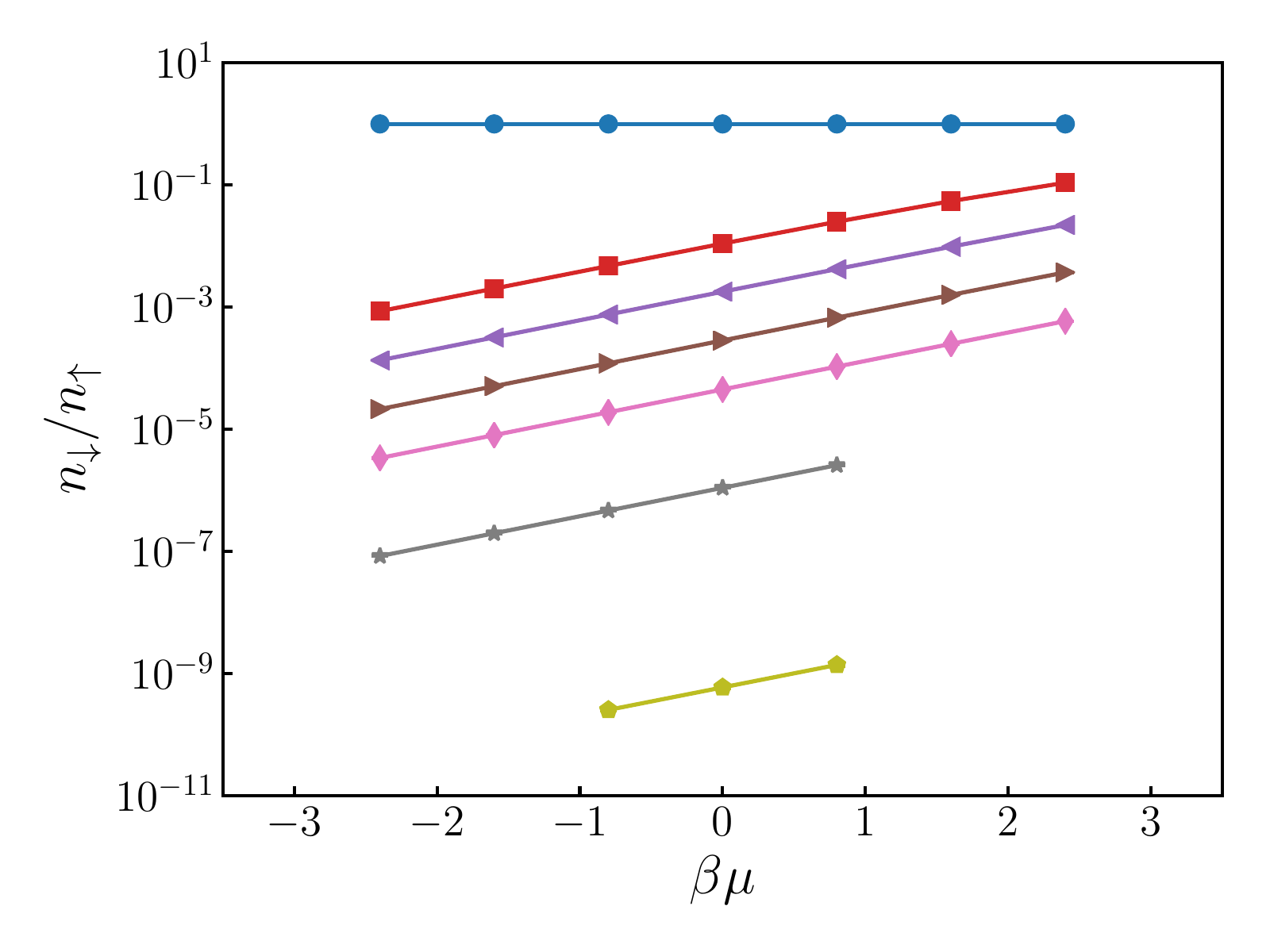}
	\caption{Dimensionless physical parameters $T/T_\text{F}$, $1/p_\text{F}a$ and the ratio $n_{\downarrow}/n_{\uparrow}$ of particle numbers
	for $N_\tau =40$ (left) and $N_\tau = 80$ (right).}
	\label{fig:particle number}
\end{figure}

In Fig.~\ref{fig:particle number},
we show dimensionless quantities $T/T_\text{F}$, $1/p_\text{F}a$ and $n_\dwn/n_\up$,
which are typical indicators of the temperature, the interaction strength 
and the population imbalance, respectively. 
The ratio of particle numbers $n_\dwn/n_\up$ becomes significantly small
when $\beta h \gg 1 \, (\beta \mu_\up \gg \beta \mu_\dwn)$ as expected.
In that case, $T/T_\text{F}$ and $1/p_\text{F}a$ are also small 
since $T_\text{F}$ and $p_\text{F}$ are proportional to $n_\up$.

\begin{figure}[hbt]
	\centering
	\includegraphics[width=0.75\linewidth]{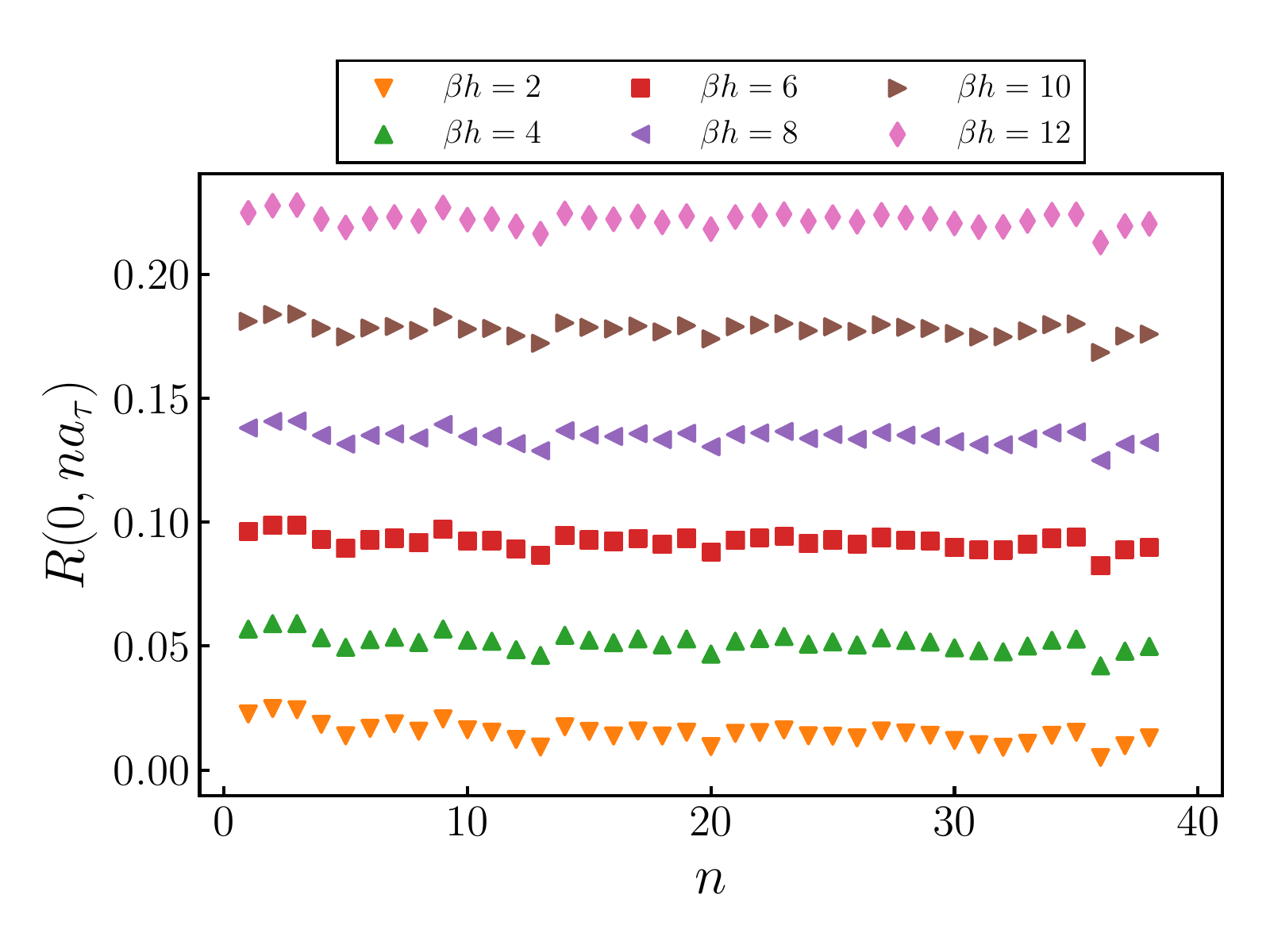}
	\caption{The $\tau$-dependence of the ratio of Green's functions
	$R(0,\tau)$ 
	on an $N_\tau = 40$ lattice at $\beta\mu=0$.
	$n$ denotes the number for the discretized imaginary time $\tau=na_\tau$. 
	Each point in this plot is obtained for a single configuration.}
	\label{fig:E0}
\end{figure}

For each parameter, 
we computed the ratio of Green's functions $R(0,na_\tau)$ 
defined in Eq.~\eqref{ratio of green functions} at zero momentum.
Numerical results on an $N_\tau = 40$ lattice at $\beta\mu=0$ 
for a single configuration are shown in Fig.~\ref{fig:E0}.
Qualitative behavior of $R(0,na_\tau)$ at other $N_\tau$ and $\beta \mu$ 
are same as these results.
In the parameter region we swept, 
$R(0,na_\tau)$ has a plateau at intermediate imaginary time,
which suggests that the energy spectrum is gapped 
from any possible excited states.
In our analysis, 
we extract the single-particle ground-state energy $E_0(0)$ by
\begin{align}
E_0(0)\simeq \frac{1}{a_\tau}R(p,\tau=(N_\tau-2)a_\tau) \label{1particle_energy}
\end{align}
because the long-time limit as 
Eq. (\ref{ratio of green functions}) 
cannot be taken on a lattice. 

\begin{figure}[hbt]
	\centering
	\includegraphics[width=0.75\linewidth]{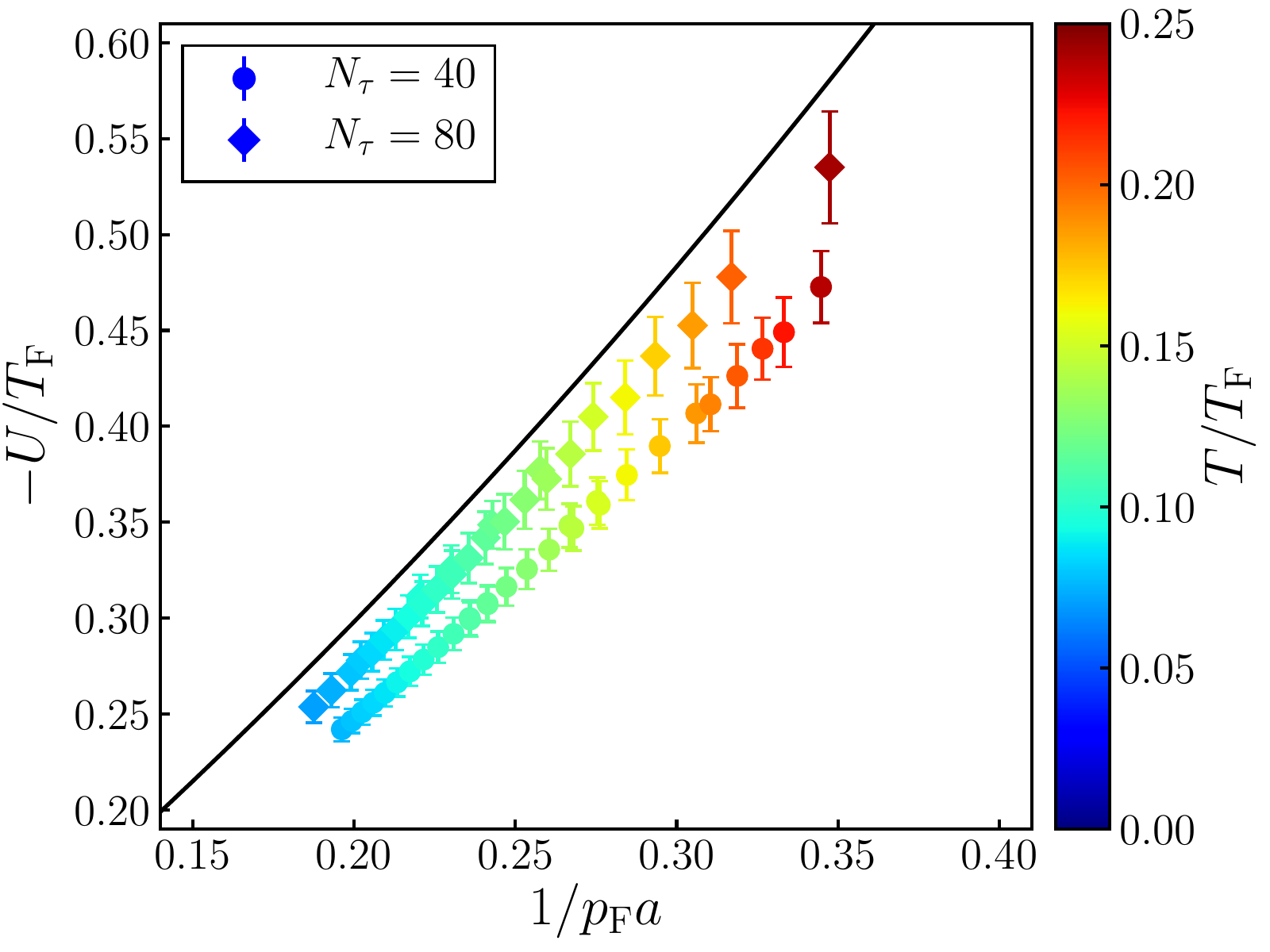}
	\caption{The polaron energy computed by the CLM.
	The solid line is the exact result at $T=0$ shown in Eq.~\eqref{exact polaron energy}.}
	\label{fig:polaron energy}
\end{figure}

After calculating the single-particle energy (\ref{1particle_energy}), 
the polaron energy $U$ is obtained by Eq. (\ref{def_polaron_energy}).
In Fig.~\ref{fig:polaron energy}, 
we show the polaron energy on $N_\tau = 40$ and $80$ lattices. 
As the temporal lattice size $N_\tau$ becomes large, the system is close to the continuum limit.
The color of each point represents the statistical average of $T/T_\text{F}$.
The lowest temperature is $T/T_\text{F} \simeq 0.08$ for $N_\tau = 40$ 
and $T/T_\text{F} \simeq 0.07$ for $N_\tau = 80$, respectively.
The ratio of particle numbers $n_\dwn/n_\up$ varies from 
$1.0 \times 10^{-5}$ to $1.5 \times 10^{-1}$ for $N_\tau = 40$ 
and 
$8.4 \times 10^{-8}$ to $1.0 \times 10^{-1}$ for $N_\tau = 80$, respectively.
The solid line indicates the exact result 
at zero temperature shown in Eq.~\eqref{exact polaron energy}.
For a fixed $N_\tau$,
the numerical results show similar behavior to Eq.~\eqref{exact polaron energy} 
as a function of $1/p_\text{F}a$
despite they also depend on $T/T_\text{F}$ and $n_\dwn/n_\up$.
Moreover, the numerical results tend to be close to the exact result at zero temperature when we take the continuum limit. Our result suggests that the polaron energy is insensitive to the temperature and the impurity concentration. 

\section{Summary}\label{sec:summary}
We have studied excitation properties of Fermi polarons at finite temperature for the attractive Gaudin-Yang model
with large population imbalances using the complex Langevin method,
a non-perturbative approach free from the sign problem.
We have performed numerical simulations 
for several chemical potential $\beta \mu$ 
and Zeeman field $\beta h$, 
the dimensionless control parameters of the model,
and found that our simulation covers wide range of 
temperature $T/T_\text{F}$, 
strength of the coupling $1/p_\text{F}a$ 
and population imbalance $n_\dwn/n_\up$.
We have computed the polaron energy as a function of $1/p_\text{F}a$.
While our result is still away from the zero temperature and single-polaron limit,
the computed polaron energy shows similar $(1/p_\text{F}a)$-dependence 
to the exact result at those limits.

The complex Langevin method well works in Gaudin-Yang model even in the presence of the population imbalance. 
Practically, within our setup, 
the probability distribution of the drift term 
always show an exponential fall-off,
which means that the problem of wrong convergence does not occur.
Moreover, the integrand of path integral is always positive within our simulation 
from the eigenvalue-analysis. 
However, it is known that 
the sign problem is severe 
in the case of higher dimension \cite{PhysRevA.82.053621}. 
Thus the behavior of the probability distribution of the drift term and
the eigenvalues in higher dimension 
will be investigated as future study.

One interesting application of the complex Langevin method is 
to study the transition from degenerate Fermi-polaron regime 
to classical Boltzmann-gas regime of a unitary spin-imbalanced Fermi gas
which is found to be a sharp transition 
by a cold-atom experiment using ${}^6$Li Fermi gases
in a three-dimensional box potential~\cite{PhysRevLett.122.093401}.
Also, it is interesting to explore an inhomogeneous pairing phase~\cite{10.21468/SciPostPhys.9.1.014,PhysRevA.103.043330} and in-medium bound states~\cite{PhysRevResearch.1.033177}, which cannot be addressed by quantum Monte Carlo simulation due to the sign problem in the mass- and population-imbalanced systems.
In order to discuss such phenomena, 
we need more elaborate estimation of systematic errors. 
The work in this direction will be presented elsewhere.

\section*{Acknowledgments}
The authors are grateful to Tetsuo Hatsuda, Kei Iida, and Yuya Tanizaki for fruitful discussion.
T.\ M.\ D. was supported by Grant-in-Aid for Early-Career Scientists (No. 20K14480).
H.\ T. was supported by Grants-in-Aid for Scientific Research from JSPS (No.\ 18H05406).
S.\ T.\ was supported by the RIKEN Special Postdoctoral Researchers Program.
This work was partly supported by RIKEN iTHEMS Program.

\bibliography{ref.bib}

\end{document}